\documentclass[prd,aps,twocolumn]{revtex4}
\usepackage{epsfig}

\begin{document}
\title{WMAP data and the curvature of space}

\author{Jean-Philippe Uzan \vskip 0.2cm}
\affiliation{Laboratoire de Physique Th{\'e}orique, CNRS--UMR
8627,
         B{\^a}t. 210, Universit{\'e} Paris XI, F--91405 Orsay Cedex,
         France,\\
         Institut d'Astrophysique de Paris, GReCO,
        CNRS-FRE 2435, 98 bis, Bd Arago, 75014 Paris, France,\\
        Department of Mathematics and Applied Mathematics,
        University of Capetown, Rondebosch 7700, Capetown, South Africa}

\author{Ulrich Kirchner and George F.R. Ellis\vskip 0.2cm}
\affiliation{Department of Mathematics and Applied Mathematics,
University of Capetown, Rondebosch 7700, Capetown, South
Africa\vskip0.1cm}

\date{February 28, 2003}

\begin{abstract}
{\it Inter alia}, the high precision WMAP data on Cosmic
Background Radiation marginally indicate that the universe has
positively curved (and hence spherical) spatial sections. In this
paper we take this data seriously and consider some of the
consequences for the background dynamics. In particular, we show
that this implies a limit to the number of $e$-foldings that could
have taken place in the inflationary epoch; however this limit is
consistent with some inflationary models that solve all the usual
cosmological problems and are consistent with standard structure
formation theory.
\end{abstract}
 \vskip2pc

\maketitle
\vskip0.5cm

The Wilkinson Microwave Anisotropy Probe (WMAP) has recently
provided high resolution Cosmic Microwave Background
data~\cite{wmap1,wmap2} of importance for cosmology - indeed they
constitute a significant contribution towards the goal of
developing precision cosmology. Among the interesting conclusions
that have been reached from this data are constraints on present
value $\Omega_0$ of the total density parameter of the universe.
The WMAP results interestingly indicate that while the universe is
close to being flat ($\Omega_0 \simeq 1$), a closed universe is
marginally preferred: $\Omega_0 > 1$~\cite{wmap1}. In particular,
with a prior on the Hubble constant, one gets that $\Omega_0 =
1.03\pm0.05$ at 95\% confidence level, while combining WMAP data
with type Ia supernovae leads to $\Omega_0 = 1.02\pm0.04$ or to
$\Omega_0 = 1.02\pm0.02$ respectively without and with a prior on
the Hubble parameter. The latter may be regarded as the present
best estimate of this parameter. Note also that this tendency to
point toward a closed universe is not strictly speaking either new
or restricted to the WMAP data. For instance with a prior on the
nature of the initial conditions, the Hubble parameter and the age
of the universe, analysis of the DASI, BOOMERanG, MAXIMA and DMR
data~\cite{sievers,netterfield} lead to $\Omega_0 = 0.99 \pm 0.12$
but to $\Omega_0 = 1.04 \pm 0.05$ if one takes into account only
the DASI, BOOMERanG and CBI data, both at $1\sigma$-level. The
Archeops balloon experiment~\cite{archeops} points toward
$\Omega_0 = 1.00_{- 0.02}^{+ 0.03}$ with data from the HST and a
prior on the Hubble constant but to values in the range $\Omega_0
= 1.16_{- 0.20}^{+ 0.24}$ from combined CBR data alone and to
$\Omega_0 = 1.04_{- 0.04}^{+ 0.02}$ from combined CBR and
supernova data, all at a 68\% confidence level. The improved
precision from WMAP is clear.

The confirmation of the existence of Doppler peaks and their
respective locations tends to confirm the inflationary
paradigm~\cite{wmapinflation}, as does the existence of an almost
scale invariant power spectrum. Furthermore a nearly flat
universe, argued to be predicted generically by most inflationary
models~\cite{linde}, is usually seen as a further evidence. Since
\begin{equation}
\delta(t)\equiv\Omega(t)-1=\frac{K}{a^2H^2}
\end{equation}
where $a(t)$ is the scale function and $H(t) = \dot{a}/a$ the
Hubble parameter, the WMAP constraints on the curvature of space
imply that the curvature density parameter, $\Omega_K(t)\equiv
K/a^2H^2$, has been smaller than a few percent from the time the
largest observable scale, $k_{\rm min}\sim a_0H_0$, crossed the
horizon during inflation up to today. It is then a good
approximation to neglect the curvature from the time $k_{\rm min}$
crossed the horizon to today when performing structure formation
calculations, mainly because $\Omega_{k_{\rm min}}=\Omega_{\rm
today}$ (see Fig.~1). The constraints on the curvature of the
universe thus back up the validity of calculations underlying the
tests of inflation performed up to now that assume a flat universe
and focus on the properties of the perturbations (spectral
indices, tensor modes, statistics, etc.).

Nevertheless, when $K \neq 0$ (which is highly favoured over the
case $K=0$ by probability considerations: the latter are of
measure zero in the space of
Friedman-Lema\^{\i}tre-Robertson-Walker universes in most
measures), $\Omega_K$ is driven toward 0 exponentially during
inflation.  The curvature term therefore was necessarily dominant
at early enough times, if inflation undergoes a sufficient number
of $e$-foldings, mainly because the curvature term behaves as
$a^{-2}$ and will tend to dominate over any kind of matter having
an equation of state less stiff than $-1/3$, and thus over a slow
rolling scalar field or a cosmological constant, in the early
inflationary era.

We argue that the study of dynamics of the background for such
models can give interesting insight into the inflationary era and
its dynamics prior to $k_{\rm min}$ horizon exit. Working backward
in time from the present, we aim to emphasize the constraints that
arise on the inflationary models from the WMAP data. In doing so,
we take seriously the indication that a value of $\Omega_0$
slightly larger than 1 is favoured. During a slow-roll
inflationary phase, the universe is described by an almost de
Sitter spacetime with the $K=+1$ scale factor
\begin{equation} a(t)\propto{\rm cosh}(Ht)
\end{equation}
instead of the exponential growth obtained in the flat case. It
follows that the number of $e$-foldings between the onset and the
end of inflation cannot be arbitrarily large, when we take present
day cosmological parameters into account. The curvature enhances
the curvature, $\ddot a$, of the scale factor in the early stage
of inflation so that there exists a turn-around point in the past.
It follows~\cite{sphericalinflation,sphericalinflation2} that the
number of $e$-foldings during inflation is bounded by a maximum
possible number of $e$-foldings, $N_{\rm max}$, related to the
present day density parameter $\Omega_0$.\

Assuming that the inflation is driven simply by a cosmological
constant, the scale factor of a closed universe in this epoch is
given by
\begin{equation}
a(t)=a_{\rm min}{\rm cosh}(\lambda t)
\end{equation}
with $\lambda\equiv\sqrt{\Lambda/3}$ and $t=0$ corresponds to the
minimum expansion of the universe. It follows that
$\delta=\Omega-1=1/{\rm sinh^2(\lambda t})$. The time $t_*$ at
which $k_{\rm min}=a_0H_0$ crosses the horizon is thus obtained to
be
\begin{equation}
\lambda t_*={\rm Arcsinh}(1/{\sqrt{\delta_0}}).
\end{equation}
The number of $e$-foldings between $t=0$ and $t_*$ is simply
\begin{equation}
\Delta N_{\rm max}=\ln\frac{a_{k_{\rm min}}}{a_{\rm min}}
\end{equation}
that is
\begin{equation}\label{6}
\Delta N_{\rm
max}=\frac{1}{2}\ln\left(1+\frac{1}{\delta_0}\right).
\end{equation}
It typically ranges from 2.3 to 1.5 when $\delta_0$ varies from
0.01 to 0.05 (see Fig.~2).

This argument also holds for negatively curved universes, in which
case, the scale factor behaves as $a\propto{\rm sinh}\lambda t$.
In that case, there will not be any maximal number of $e$-foldings
but the universe starts off from a singularity. This will enhance
the trans-Planckian problem~\cite{transplanck}.\\

Another way to look at the existence of a maximum number of
$e$-foldings in the $K=+1$ case, is to relate the number of
$e$-foldings during inflation to the later history of the
universe~\cite{sphericalinflation,sphericalinflation2}. During
inflation, the comoving Hubble length $aH$ is decreasing, while it
is increasing at all further times. A given comoving scale, $k$,
crosses the horizon when $k=a(t_k)H(t_k)\equiv a_kH_k$. The number
of $e$-foldings, $N(k)$, between when this scale crosses the
horizon and the end of inflation is obtained to be
\begin{equation}
N(k)=\ln\left(\frac{a_{\rm end}H_{\rm end}}{a_kH_k}\right).
\end{equation}
The determination of $N(k)$ requires that we compute
\begin{equation}
{\cal R}\equiv\frac{a_0}{a_{\rm end}}
\end{equation}
which requires a complete history of the evolution of the universe
from the end of inflation to today. If one assumes that the
universe is matter dominated from the end of (slow roll) inflation
to reheating, radiation dominated from reheating to equality and
matter dominated up to now~\footnote{The results will be only
marginally affected by a late-time cosmological constant as
indicated by the supernova data.}, then~\cite{lythliddle,lidsey}
\begin{eqnarray}
N(k)&\simeq&62-\ln\left(\frac{k}{a_0H_0}\right)
-\ln\left(\frac{10^{16}\,{\rm GeV}}{V_{k}^{1/4}}\right)\nonumber\\
&&+\frac{1}{4}\ln\frac{V_k}{V_{\rm
end}}-\frac{1}{3}\ln\left(\frac{V_{\rm end}^{1/4}}{\rho_{\rm
reh}^{1/4}}\right)-\ln h,
\end{eqnarray}
$h$ being the Hubble parameter in unites of 100~km.s$^{-1}$/Mpc.
For our purpose, it will be more convenient to relate $N(k)$ to
${\cal R}$ which can easily be found to be given by
\begin{equation}
-\ln{\cal R}\simeq-66-\frac{1}{3}\ln\left(\frac{V_{\rm
end}^{1/4}}{\rho_{\rm
reh}^{1/4}}\right)+\ln\left(\frac{10^{16}\,{\rm GeV}}{V_{\rm end
}^{1/4}}\right).
\end{equation}
It follows that
\begin{eqnarray}
N(k)&\simeq&128-\ln{\cal R}-\ln\left(\frac{k}{a_0H_0}\right)
\nonumber\\
&&-2\ln\left(\frac{10^{16}\,{\rm GeV}}{V_{k}^{1/4}}\right)-\ln h.
\end{eqnarray}

The cosmic microwave background roughly probe scales from 10~Mpc
(i.e. of the order of the thickness of the last scattering
surface) to $10^4$~Mpc (i.e. of the order of the size of the
observable universe) while galaxy surveys can probe the range
1~Mpc to 100~Mpc.

Concerning the characteristic scales involved in the problem, we
can assume that $V_k\sim V_{\rm end}$ as long as slow rolling
holds. The reheating temperature can be argued to be larger than
$\rho_{\rm reh}^{1/4}>10^{10}$~GeV to avoid the gravitino
problem~\cite{sarkar} and may be pushed in the extreme case to
$10^3$~GeV, i.e. just before the electro-weak transition so, that
baryogenesis can take place. The amplitude of the cosmological
fluctuations (typically of order $2\times10^{-5}$ on Hubble
scales) roughly implies that $V_{\rm end}^{1/4}$ is smaller than a
few times $10^{16}$~GeV and, for the same reason as above, has to
be larger than $10^{10}$~GeV in the extreme case.

This implies that the number of $e$-foldings has approximately to
lie between 50 and 70, which is the order of magnitude also
required to solve the horizon and flatness problem when
$K=0$~\cite{linde}, but it can be lowered to 25 in the extreme
case of thermal inflation~\cite{thermal}, which we are not
considering here.

In the following, we restrict our analysis to the case where
reheating takes place just at the end of inflation, i.e.
$\rho_{\rm reh}=V_{\rm end}$. It has been shown
\cite{sphericalinflation} that, in that case, the existence of a
positive curvature limits the number of $e$-foldings to
\begin{equation}\label{bound}
N_{\rm max}({\cal R}, \delta_0)=\ln\sqrt{\frac{\alpha}{\delta_0}}
\end{equation}
with
\begin{equation}
\alpha=(1+\delta_0-\Omega_{\rm rad0}){\cal R}+\Omega_{\rm
rad0}{\cal R}^2
\end{equation}
where $\delta_0 \equiv \Omega_0 -1$ and $\Omega_{\rm
rad0}\simeq4.17\times10^{-5}h^{-2}$ is the radiation density
parameter today. Let us emphasize that the derivation of
Eq.~(\ref{bound}), as well as Eq.~(\ref{6}), does not make any
assumption concerning what happened before the start of inflation.

This allows us to estimate the number of $e$-foldings allowed
prior to the time at which $k_{\rm min}$ crossed the horizon,
\begin{equation}
\Delta N=N_{\rm max}[{\cal R}(V_{\rm end}^{1/4}),
\delta_0]-N(k_{\rm min},V_{\rm end}^{1/4})
\end{equation}
where we will let $\delta_0$ vary in the range $0.01-0.05$ (see
Fig.~2). It can be checked, as expected from Eq.~(\ref{bound})
that $\Delta N$ depends slightly on $V_{\rm end}^{1/4}$.

We see that the allowed number of $e$-foldings are compatible with
the requirements of structure formation. `Flatness' is of course
solved to the accuracy represented by $\delta_0$. The  scenarios
sketched here do not lead to as small a value of $\delta_0$ as is
often supposed -  and thereby is compatible with the best-fit WMAP
data. However, for the mean value of the curvature, one obtains
that
\begin{equation}\label{15}
\delta_0=0.02\Longleftrightarrow\Delta N_{\rm max}=1.97.
\end{equation}
With standard parameter values this gives the maximum possible
number of $e$-foldings as about $N_{\rm max}=62+1.97 \simeq 64$ -
compatible with the estimates given above, but not nearly as large
as suggested in some models of inflation.

\begin{figure}
\centerline{\psfig{figure=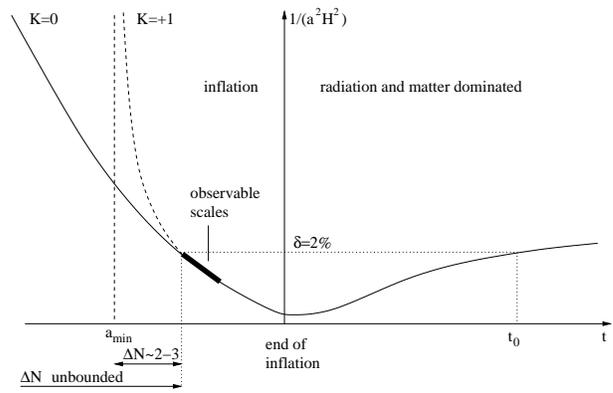,width=8cm}} \vskip0.2cm
\caption{The curvature ($1/a^2H^2$) as a function of time
(schematically). The solid line represents the case of a flat
universe ($K=0$)with exponential inflation while the dashed line
depicts the case of a closed universe ($K=+1$). Before the end of
inflation $aH$ scales either as $\hbox{e}^{\lambda t}$ ($K=0$) or
$1/{\rm sinh}\lambda t$ ($K=+1$) while it scales as $\alpha
t^{\alpha-1}$ during matter and radiation dominated eras. The
largest observed scale is depicted as well, and one can easily see
that $\Omega_K$ will be smaller than a few percent between the
time it crosses the horizon during inflation to today when it
reenters the horizon.} \label{fig1}
\end{figure}

\begin{figure}
\centerline{\epsfig{figure=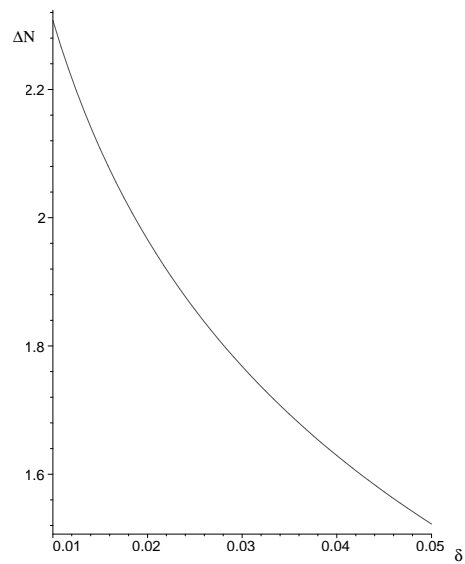,width=6cm}} \vskip0.2cm
\caption{The maximum number of $e$-foldings allowed before the
largest observable scale, $k_{\rm min}$ exits the horizon during
inflation, as a function of $\delta_0$.} \label{fig2}
\end{figure}

In view of the bounds on $N_{\rm max}$, there are also bounds on the
ability of inflation to solve the horizon problem when one considers
inflationary dynamics with $K=+1$ \cite{sphericalinflation2}. Roughly
speaking, the horizon problem is only solved during the inflationary
era itself if the start of the inflationary era is close enough to the
throat at $t=0$ in the expansion (see Eq.~3); that is, if the total
number $N$ of $e$-foldings during inflation is close to the total
$N_{\rm max}$ allowed, i.e. $N\simeq N_{\rm max}$. If inflation starts
sufficiently far from that throat, then it does not matter how many
$e$-foldings take place, they cannot provide causal connectivity on
large enough scale to solve the horizon problem; rather that
connectivity has to be set up in the pre-inflationary era (and is
evidenced by a smooth enough small patch that is then expanded to
large scales by inflation). However inflation certainly can start near
enough the throat to solve the horizon problem during the inflationary
era and still lead to $\delta_0 = 0.02$. Examples of non-slow rolling
cases are the emergent and awakening universes~\cite{emergent} that
start form an Einstein static state and are singularity free.  The
above formula (\ref{15}) limiting the number of $e$-foldins will not
apply in these cases.

In this note we have argued that the curvature of space may play
an important role in the dynamics of inflation in its early stage.
We have argued that a maximum number of 1.5 -- 2.5 $e$-foldings
can take place before the largest scale observed in the universe
crossed the horizon during slow-roll inflation. However, despite
the fact that the curvature term will be smaller than a few
percent during all times ranging from the horizon exit of the
largest observable mode to its reentry today, a positive curvature
term as small as the one indicated by current data will
dramatically change the early phase of inflation and hence lead to
the limits on the number of $e$-foldings allowed as described
above. These restrict the families of inflationary universes
compatible with the current data, but do not exclude inflationary
models even though the value of $\delta_0$ indicated by WMAP is
larger than supposed in most inflationary scenarios.

Note a possible way around the argument. In the case that
inflation is driven by a scalar field, the existence of a maximal
number of $e$-foldings may be avoided if it exits the slow-roll
regime. It was however shown that the dynamics of closed universes
filled with a scalar field is chaotic~\cite{page,topo2,cornish}
and can lead to the existence of singularity free
universes~\cite{pfs} with periodic~\cite{swh} or
aperiodic~\cite{page} trajectories. In that case, again, a
positive curvature affects the dynamics of the early phase of
inflation and the slow-roll regime has to be left before $N_{\rm
max}$ is reached, which is also at odds with the standard picture.

Let us finally consider further features of spherical universes
that may turn to out be useful in understanding the WMAP data.
Surprisingly, the WMAP angular correlation function seems to lack
signal strength on angular scales larger than 60
degrees~\cite{wmap1,wmap2}. This may indicate a possible
discreteness of the initial power spectrum, as expected e.g. from
non-standard topology. The case for topology becomes most
interesting in the case of spherical spaces~\cite{gaus}. First,
the spectrum of the Laplacian in spherical spaces is always
discrete~\cite{topomodes,topomap}. Second, and as emphasized in
Ref.~\cite{topo}, a non trivial topology is most likely to be
detectable in the case of spherical spaces and examples of the
resulting observational effects are discussed in
Ref.~\cite{topospherique}.

The same conclusion cannot be reached for negatively curved
universe. It is unlikely that we could detect a compact hyperbolic
space because, for an observer at a generic point, the topology
scale is comparable to the curvature radius or longer. The
possibility of a detection would require the observer to sit near
a short closed geodesic~\cite{weeks}.

In conclusion, if the indication that $\Omega_0$ differs from
unity at a level of a few percent is confirmed then it may imply
important results concerning the dynamics of the early stage of
our universe. The $K=+1$ background de Sitter model (3) differs
fundamentally from the scale-free de Sitter model in the $K=0$
frame - not least in terms of being geodesically complete. The
WMAP observation suggest the former may be the appropriate model
to use in investigating early inflationary dynamics.

\vskip0.25cm
{\bf Acknowledgements:} JPU thanks the Department of
Mathematics and Applied Mathematics of the University of Cape Town
for hospitality. UK acknowledges financial support from the
University of Cape Town. We thank Roy Maartens for his comments.



\begin{references}
\bibitem{wmap1} C.L. Bennett {\em et al.}, [{\tt
arXiv:astro-ph/0302207}].

\bibitem{wmap2} D.N. Spergel {\em et al.}, [{\tt
arXiv:astro-ph/0302209}].

\bibitem{sievers}
J. L. Sievers {\em et al.}, [{\tt arXiv:astro-ph/0205387}]

\bibitem{netterfield}
C.B.~Netterfield {\em et al.}, Astrophys. J. {\bf 571}, 604
(2002).

\bibitem{archeops}
A. Beno\^\i{}t {\em et al.}, [{\tt arXiv:astro-ph/0210306}]; A.
Beno\^\i{}t {\em et al.}, [{\tt arXiv:astro-ph/0210305}].

\bibitem{wmapinflation}
H.V. Peiris {\em et al.}, [{\tt arXiv:astro-ph/0302225}]; V.
Barger, H.S. Lee, and D. Marfatia, [{\tt arXiv:hep-ph/0302150}].

\bibitem{linde}
A.H.~Guth, Phys. Rev. D {\bf 23}, 347 (1981); A.D.~Linde, Phys.
Lett. B {\bf 108}, 389 (1982); S.W.~Hawking, Phys. Lett. B {\bf
115}, 295 (1982); A.A.~Starobinski, Phys. Lett. B {\bf 117}, 175
(1982); A.~Albrecht and P.~J.~Steinhardt, Phys. Rev. Lett. {\bf
48}, 1220 (1982).

\bibitem{sphericalinflation}
G.F.R. Ellis {\em et al.}, Gen. Rel. Grav. {\bf 34}, 1445 (2002).

\bibitem{sphericalinflation2}
G.F.R. Ellis {\em et al.}, Gen. Rel. Grav. {\bf 34}, 1461 (2002).

\bibitem{transplanck}
J. Martin and R. Brandenberger, Phys. Rev. D {\bf65}, 123501
(2001); R. Brandenberger and J. Martin, Mod. Phys. Lett. A
{\bf16}, 999 (2001); M . Lemoine {\em et al.}, Phys. Rev. D {\bf
65}, 023510 (2002).

\bibitem{lythliddle}
D. Lyth and A. Liddle, Phys. Rept. {\bf 231}, 1 (1993).

\bibitem{lidsey}
J.E. Lidsey {\em et al.}, Rev. Mod. Phys. {\bf69}, 373 (1997).

\bibitem{sarkar}
S. Sarkar, Rep. Prog. Phys. {\bf 59}, 1493 (1996).

\bibitem{thermal}
D.H. Lyth and E.D. Stewart, Phys. Rev. D {\bf53}, 1784 (1996);
{\em ibid.}, Phys. Rev. Lett. {\bf75}, 201 (1995).

\bibitem{emergent}
G.F.R. Ellis and R. Maartens, [{\tt arXiv:gr-qc/0211082}].

\bibitem{page} D.~N.~Page, Class. Quant. Grav. {\bf 1}, 417 (1984).

\bibitem{topo2} A.~V.~Toporensky, [{\tt arXiv:gr-qc/0011113}].

\bibitem{cornish} N.~Cornish and E.~P.~S. Shellard,
Phys. Rev. Lett. {\bf 81}, 3571 (1998).

\bibitem{pfs} L.~Parker and S.~A.~Fulling, Phys. Rev. D {\bf 7}, 2357
(1973); A.~A.~Starobinsky, Sov. Astron. Lett. {\bf 4},82 (1978).

\bibitem{swh} S.~W.~Hawking, in {\it Relativity, Group and Topology
II}, edited by B.~S.~de~Witt and R.~Stora, (North Holland,
Amsterdam, 1984).

\bibitem{gaus}
E.~Gausmann {\em et al}., Class.\ Quant.\ Grav.\ {\bf 18}, 5155
(2001).

\bibitem{topomodes}
R. Lehoucq, J.-P. Uzan, and J. Weeks, Kodai Math. J. (to appear),
[{\tt arXiv:math.SP/0202072}]; R. Lehoucq {\em et al.}, Class.
Quant. Grav. {\bf 19}, 4683 (2002).

\bibitem{topomap}
A. Riazuelo {\em et al.}, [{\tt arXiv:astro-ph/0212223}].

\bibitem{topo}
J. Weeks, R. Lehoucq, and J.-P. Uzan, Class. Quant. Grav. (to
appear), [{\tt arXiv:astro-ph/0209389}].

\bibitem{topospherique}
A. Riazuelo {\em et al.}, ``Cosmic microwave background maps of
multi-connected spherical spaces" (in preparation).

\bibitem{weeks}
J.~Weeks, Int. J. Mod. Phys. A (submitted); G.I.~Gomero,
M.J.~Rebou{\c c}as, and R.~Tavakol, Class.\ Quant.\ Grav.\ {\bf
18}, 4461 (2001); G.I. Gomero, M.J. Rebou{\c c}as, and R. Tavakol,
Int. J. Mod. Phys. A {\bf 17}, 4261 (2002).

\end{references}
\end{document}